\pdfoutput=1
\documentclass[11pt,letterpaper]{article}

\usepackage[T1]{fontenc}
\usepackage[utf8]{inputenc}
\usepackage{lmodern}
\usepackage[margin=1in]{geometry}
\usepackage{amsmath,amssymb,amsthm,mathtools}
\usepackage{microtype}
\usepackage{enumitem}
\usepackage{algorithm}
\usepackage{algpseudocode}
\usepackage{xcolor}
\usepackage{tikz}
\usetikzlibrary{arrows.meta,calc,positioning}
\usepackage[numbers,sort&compress]{natbib}
\usepackage[
  colorlinks=true,
  linkcolor=blue!45!black,
  citecolor=blue!45!black,
  urlcolor=blue!45!black,
  pdftitle={A Deterministic O(log m)-Competitive Algorithm for Online Minimum Cost Bipartite Matching with Delays}
]{hyperref}
\newcommand{\doi}[1]{doi:\,\href{https://doi.org/#1}{\nolinkurl{#1}}}

\newtheorem{theorem}{Theorem}[section]
\newtheorem{lemma}[theorem]{Lemma}

\theoremstyle{definition}
\newtheorem{definition}[theorem]{Definition}
\theoremstyle{remark}

\newcommand{\ALG}{\operatorname{ALG}}
\newcommand{\OPT}{\operatorname{OPT}}
\newcommand{\RM}{\textnormal{RM}}
\newcommand{\age}{\operatorname{age}}
\newcommand{\Rfree}{R_{\mathrm{free}}}
\newcommand{\Moff}{M^{\mathrm{OFF}}}
\newcommand{\Mout}{M^{\mathrm{VRM}}}
\newcommand{\symdiff}{\mathbin{\triangle}}
\DeclareMathOperator{\ter}{ter}

\providecommand{\Description}[2][]{}
\newenvironment{widefigure}[1][tbp]{\begin{figure}[#1]}{\end{figure}}

\numberwithin{equation}{section}
\setlist[enumerate]{itemsep=3pt,topsep=5pt}
\setlist[itemize]{itemsep=3pt,topsep=5pt}
\allowdisplaybreaks[1]

\title{A Deterministic $O(\log m)$-Competitive Algorithm for\\
Online Minimum Cost Bipartite Matching with Delays}
\author{Tung-Wei Kuo\\[2pt]
\normalsize Department of Computer Science, National Chengchi University, Taipei, Taiwan\\
\normalsize \texttt{twkuo@cs.nccu.edu.tw}}
\date{}

\begin{document}
\maketitle

\begin{abstract}
In the online Minimum cost Bipartite Perfect Matching with Delays (MBPMD)
problem, $m$ requests and $m$ servers arrive over time in a metric space.
An online algorithm has to match each request to a server and can postpone
a matching decision by paying the delay cost. The objective is to minimize
the total distance and delay cost. Azar, Chiplunkar, and Kaplan gave a
randomized $O(\log n)$-competitive algorithm for metric spaces with $n$
points that are known in advance. For deterministic algorithms, however,
the best known competitive ratio in general metric spaces was
$O(m^{0.59})$, due to Azar and Jacob-Fanani. On the line, Kuo proposed the
Virtual Robust Matching (VRM) algorithm and showed that it is
$O(\sqrt{m}\log^2 m)$-competitive.

We prove that VRM is $O(\log m)$-competitive for the MBPMD problem in any
metric space. VRM is deterministic and does not need to know the metric
space or $m$ in advance. It is the first deterministic
polylogarithmic-competitive algorithm for the MBPMD problem in general
metric spaces. Unlike the previous analysis of VRM, our analysis does not
rely on a geometric decomposition of the metric space. Instead, it
charges the waiting time of requests over time to the edges of a fixed
optimal matching.

We also extend VRM to the non-bipartite version and prove that the
resulting deterministic algorithm is $O(\log m)$-competitive. This
improves the $O(\log^5 m)$ bound of Dufay and Wattenhofer and matches a
concurrent result of Ahmadian, Chawla, Kumar, Purohit, and Zhang.

The proof that VRM is $O(\log m)$-competitive and its non-bipartite
extension were found by GPT-6 Astra (OpenAI) with almost no human input. We used Claude Opus 5.5 (Anthropic)
to simplify and present the results. See the AI Usage and Methodology
section for details. We take full responsibility for this paper.

\end{abstract}

\section{Introduction}
\label{sec:intro}

Consider a platform that matches passengers to drivers. When a passenger
arrives, the platform can assign an available driver immediately or wait
for a driver who is closer to the passenger. Waiting may reduce the
distance traveled, but it also increases the waiting time of passengers
and drivers. Thus, a natural objective is to minimize the sum of the
distance and waiting time.

The above tradeoff is captured by the Minimum cost Bipartite Perfect
Matching with Delays (MBPMD)
problem~\cite{azar2016polylogarithmic,ashlagi2017min}. In this problem,
$m$ requests and $m$ servers arrive over time in a metric space. An
online algorithm has to match each request to a server, and can
postpone a matching by paying the delay cost. Once a request and a server
are matched, this decision cannot be changed. The objective is to minimize
the total distance and delay cost. The MBPMD problem is the bipartite
variant of the Minimum cost Perfect Matching with Delays (MPMD)
problem~\cite{emek2016online}, in which any two requests can be matched
together.

We analyze algorithms using the standard notion of competitive ratio.
Specifically, an online algorithm is $c$-competitive if, for any
input, its cost is at most $c$ times the cost of the optimal offline
algorithm.

For the MBPMD problem, Azar, Chiplunkar, and Kaplan proposed a randomized
$O(\log n)$-competitive algorithm, where $n$ is the number of points in the
metric space, and proved that any randomized algorithm has competitive
ratio $\Omega(\log^{1/3}n)$~\cite{azar2016polylogarithmic}.
This lower bound was improved to
$\Omega\bigl(\sqrt{\log n/\log\log n}\bigr)$~\cite{ashlagi2017min}. Both
lower-bound constructions place $\Theta(n)$ requests on a line, so the
lower bound $\Omega\bigl(\sqrt{\log m/\log\log m}\bigr)$ holds even on a
line whose points are known in advance.

For the MPMD problem, Emek, Kutten, and Wattenhofer proposed a randomized
algorithm with competitive ratio $O(\log^2 n+\log\Delta)$, where $\Delta$
is the aspect ratio of the metric space~\cite{emek2016online}. Azar,
Chiplunkar, and Kaplan improved this ratio to $O(\log n)$ and proved an
$\Omega(\sqrt{\log n})$ lower bound for randomized
algorithms~\cite{azar2017polylogarithmic}. The same paper that improved
the MBPMD lower bound also improved this one to
$\Omega(\log n/\log\log n)$~\cite{ashlagi2017min}.

The known randomized algorithms first embed the metric space into a distribution over Hierarchically
Separated Trees (HSTs)~\cite{fakcharoenphol2003tight}, so they need to know
the metric space in advance. Recently, Coester and Huang proposed
probabilistic online monotone embeddings into HSTs, which do not require
prior knowledge of the metric space, but no competitive ratio for the MPMD
or MBPMD problem has been derived from them, and deterministic online
monotone embeddings have linear distortion~\cite{coester2026online}.

In contrast, several deterministic algorithms do not require prior
knowledge of the metric space. For the MPMD problem, Bienkowski, Kraska,
and Schmidt proposed an $O(m^{2.46})$-competitive
algorithm~\cite{bienkowski2017match}. Bienkowski, Kraska, Liu, and
Schmidt improved this ratio to $O(m)$~\cite{bienkowski2018primal}, and
Azar and Jacob-Fanani obtained $O(\epsilon^{-1}m^{\log_2(3/2+\epsilon)})$
for any fixed $\epsilon>0$, which is $O(m^{0.59})$ for small enough
$\epsilon$~\cite{azar2020deterministic}. Both groups adapted their
algorithms to the MBPMD problem with the same competitive ratios, and the
analyses are tight already on metric spaces with two points and one
point, respectively.

Kuo proposed the Virtual Robust Matching (VRM) algorithm for the MBPMD
problem and proved an $O(\sqrt m\log^2 m)$ competitive ratio on a
line~\cite{kuo2025online}. Very recently, Dufay and Wattenhofer improved
the deterministic ratio for the MPMD problem from polynomial to
polylogarithmic in $m$. Their $O(\log^5 m)$-competitive algorithm does
not need to know the metric space or $m$ in
advance~\cite{dufay2026deterministic}. However, they noted that several
key parts of their approach, specifically the greedy algorithm and the
component decomposition, do not naturally extend to the bipartite
case~\cite{dufay2026deterministic}.

Concurrently with and independently of our work, Ahmadian, Chawla, Kumar,
Purohit, and Zhang
obtained a deterministic $O(\log m)$-competitive algorithm for the MPMD
problem that also does not need to know the metric space or $m$ in
advance. They noted that their approach does not extend directly to the
bipartite case and left a logarithmic competitive ratio for this case as
an open question~\cite{ahmadian2026general}. Thus, prior to our work, no
polylogarithmic competitive ratio was known for the MBPMD problem without
prior knowledge of the metric space.

\subsection{Our Contribution and Techniques}
\label{sec:contribution}

In this paper, we prove that the VRM algorithm is $O(\log m)$-competitive
for the MBPMD problem in any metric space. Specifically, let
$\ALG$ be the cost of the online algorithm and let $\OPT$ be the cost of
the optimal offline algorithm, which knows the entire input. We have the
following result.

\begin{theorem}
\label{thm:main}
For every fixed $\gamma>1$, the VRM algorithm with parameter $\gamma$
described in Section~\ref{sec:algorithm} is a deterministic online algorithm for the
MBPMD problem in any metric space, and on every input, its cost
satisfies
\[
  \ALG\le
  \left(2+\frac{2\gamma}{\gamma-1}\right)
  \bigl[1+(\gamma+1)(1+\ln m)\bigr]\OPT.
\]
In particular, for $\gamma=2$, the VRM algorithm is $(24+18\ln m)$-competitive.
Moreover, the algorithm does not need to know the metric space or $m$
in advance.
\end{theorem}

Theorem~\ref{thm:main} gives the first deterministic algorithm with a
polylogarithmic competitive ratio for the MBPMD problem in general metric
spaces. In particular, it gives a logarithmic competitive ratio for the
bipartite problem, which~\cite{ahmadian2026general} leaves open.

\paragraph{The RM algorithm.}
VRM is based on the Robust Matching (RM) algorithm proposed by Raghvendra
for the online Minimum cost Bipartite Perfect Matching (MBPM)
problem~\cite{raghvendra2016robust}. In the MBPM problem, all servers
arrive in the beginning, and a request must be matched immediately after
it arrives. The objective is the total distance of the matching.

The RM algorithm maintains an offline matching $\Moff$ and an online matching
$M^{\RM}$, which is the real output matching. When a request $r$ arrives,
the algorithm computes an $\Moff$-augmenting path $P_r$ from $r$ to a free
server $s$ that minimizes the \emph{$\gamma$-net-cost}, which is $\gamma$
times the total length of the edges added to $\Moff$ minus the total
length of the edges removed from $\Moff$.\footnote{In~\cite{raghvendra2016robust},
this cost is referred to as the $t$-net-cost. However, because $t$
denotes time in this paper, we change $t$-net-cost to $\gamma$-net-cost.}
It then adds the pair $rs$ to
$M^{\RM}$ and augments $\Moff$ by $P_r$. Nayyar and Raghvendra proved that
the competitive ratio of RM is $O(m^{1-1/d}\log^2 m)$ in $d$-dimensional
Euclidean space~\cite{nayyar2017input}, and Raghvendra proved that
RM is $O(\log m)$-competitive on a line~\cite{raghvendra2018optimal}.

\paragraph{The VRM algorithm.}
The MBPMD problem differs from the MBPM problem in two respects: servers
arrive over time, and a request can wait. To handle the arrival times, VRM
works in the \emph{time-augmented} (TA) metric. In the TA metric, the
distance between two items (called their \emph{TA distance}) is their
distance in the original metric space plus the difference of their
arrival times~\cite{bienkowski2017match,azar2020deterministic,kuo2025online}.

To decide when to match a request, the original VRM algorithm introduces
a \emph{Moving Virtual (MV) server} for every request. The MV server of a
request stays at the location of the request and moves forward in time,
so its TA distance to the request is the time elapsed since the request
arrived. Consider a request $r$. An augmenting path of $r$ is
\emph{real} if all servers in the path are real. It is \emph{virtual} if
its last edge is from some request $p$ to $p$'s MV server, and all
the other servers in the path are real.

For example, the edge from $r$ to $r$'s MV server is a virtual augmenting
path of $r$. Its $\gamma$-net-cost is $\gamma$ times the waiting time of
$r$, so it is zero when $r$ arrives and increases over time. Among the real
augmenting paths of $r$, let $P_r$ be one with minimum $\gamma$-net-cost.
Define $\widetilde{P}_r$ similarly for the virtual augmenting paths of $r$.
The request $r$ is matched when the $\gamma$-net-cost of $P_r$ becomes
less than or equal to that of $\widetilde{P}_r$.

Kuo proved that the edge from $r$ to $r$'s MV server always has the
minimum $\gamma$-net-cost among the virtual augmenting paths of
$r$~\cite[Lemma~4.1]{kuo2025online}.\footnote{The lemma is
stated there for $\gamma=3$ on the line, but its proof needs only
$\gamma\ge1$ and does not use the line structure. Hence, it holds for
every $\gamma>1$ in every metric space.} Thus, $r$ is
matched when the
$\gamma$-net-cost of $P_r$ becomes less than or equal to $\gamma$ times the
waiting time of $r$. In Section~\ref{sec:algorithm}, we present this
simplified algorithm, which we take as the definition of VRM in this
paper.

\paragraph{The analysis.}
Let $W_R$ be the total waiting time of requests under VRM. Our analysis
has two steps. The first step bounds the total cost of the algorithm,
including the distance cost and the delay cost of both requests and
servers, by $(2+\frac{2\gamma}{\gamma-1})W_R$. The proof of this step is
similar to the analysis of VRM~\cite{kuo2025online}.

The second step is a charging argument, in which we charge $W_R$ to the
optimal cost. We fix an optimal matching $M^*$. At any time $t$, the
symmetric difference of $M^*$ and $\Moff$ contains an alternating path from
each free request $r$ to a server. We call this path the
\emph{optimal path} of $r$ and the server its \emph{terminal server}.
At every moment, we charge the waiting cost of $r$ to the edges of $M^*$ on
its optimal path. There are two cases, according to whether the terminal
server of the path has arrived.
\begin{enumerate}
\item If the terminal server $s$ has not arrived, we charge the waiting
cost of $r$ to the last edge $e=ps$ of the path, where $p$ is the
request preceding $s$ on the path. The optimal path is an
$\Moff$-augmenting path, so $e$ is not in $\Moff$ and hence is an edge of
$M^*$. Because the optimal paths of distinct free requests share no edge,
$e$ is charged by at most one request at any time. Moreover, $e$ is charged only after $p$ arrives and
before $s$ arrives. Hence, the total charge to $e$ is at most the
difference of the arrival times of $p$ and $s$, and summing over the edges
of $M^*$ shows that the total charge in this case is at most $\OPT$.
\item If the terminal server has arrived, then the optimal path is a valid
augmenting path that the algorithm could use. Because $r$ is waiting, the
$\gamma$-net-cost of this path is at least $\gamma$ times the waiting time
of $r$. Consequently, the waiting time of $r$ is at most the total length
$D^*$ of the edges of $M^*$ on the path, and every vertex on the path arrived
within the last $(\gamma+1)D^*$ time units. We charge the waiting cost of $r$
to the edges of $M^*$ on the path, in proportion to their lengths.
Each edge $e$ of $M^*$, of length $D(e)$, receives a total charge of at most
$(\gamma+1)({\OPT}/m+D(e)\ln m)$ in this case.
\end{enumerate}
Summing over the $m$ edges of $M^*$, the total charge in the second case
is at most $(\gamma+1)(1+\ln m)\OPT$. Thus,
$W_R\le[1+(\gamma+1)(1+\ln m)]\OPT$, and Theorem~\ref{thm:main} follows
from the first step.

\paragraph{Extension to the MPMD problem.}
In Section~\ref{sec:nonbipartite}, we extend VRM and its analysis to the
MPMD problem. We call the resulting algorithm non-bipartite VRM. By
Theorem~\ref{thm:nonbipartite}, it is $(28+24\ln m)$-competitive for
$\gamma=2$, where $2m$ is the number of requests. This improves the
$O(\log^5 m)$ bound of~\cite{dufay2026deterministic} to $O(\log m)$.
Like VRM, it is deterministic and does not need to know the metric space
or $m$ in advance. In concurrent work, Ahmadian, Chawla, Kumar, Purohit,
and Zhang obtained the same asymptotic bound~\cite{ahmadian2026general}.

The extension requires two main changes. First, because there are no
servers, both endpoints of an augmenting path are requests, and each of
them can initiate an augmentation. Thus, both endpoints of an optimal path
can be free requests that charge the edges of $M^*$ on the path at the
same time, which at most doubles the charge rate of each such edge.
Second, in the bipartite case, VRM computes the minimum augmenting path
$P_r$ of each request $r$ as a shortest path in a directed graph whose
orientation relies on the bipartition. In the
non-bipartite case, we instead compute $P_r$ from minimum-weight perfect
matchings in general graphs.

\subsection{Other Related Work}
\label{sec:related}

\paragraph{Special metric spaces.}
When the metric space is known in advance and forms a tree of height $h$,
deterministic $O(h)$-competitive algorithms are known for both the MPMD
and the MBPMD
problems~\cite{azar2017polylogarithmic,azar2016polylogarithmic,ashlagi2017min}.
On a
line with $n$ points, however, this gives only an $O(n)$ bound. When
$n=2$, Emek, Shapiro, and Wang proposed a deterministic $3$-competitive
algorithm for the MPMD problem and proved that $3$ is optimal for
deterministic algorithms~\cite{emek2019minimum}. For randomized
algorithms, a $2$-competitive algorithm is known for this case, and it is
optimal~\cite{he2023randomized}.

\paragraph{Online metric matching without delays.}
VRM is based on the RM algorithm for the MBPM problem. For this problem,
the optimal deterministic competitive ratio in general metric spaces is
$2m-1$~\cite{kalyanasundaram1993online,khuller1994line}. Randomized
algorithms achieve $O(\log^2 m)$ in general~\cite{bansal2014randomized}
and $O(\log m)$ for doubling metrics~\cite{gupta2012online}. Nayyar and
Raghvendra showed that the RM algorithm is
$O(m^{1-1/d}\log^2 m)$-competitive when the servers span a space of
doubling dimension $d$~\cite{nayyar2017input}, and Raghvendra showed that
it is $O(\log m)$-competitive on a line~\cite{raghvendra2018optimal}. No
randomized algorithm on a line is
$o(\sqrt{\log m})$-competitive~\cite{peserico2023matching}. In contrast,
no algorithm has a bounded competitive ratio for a non-bipartite variant,
even on a line under random-order arrivals~\cite{ascher2026online}. In
this variant, all requests arrive online, and each arriving request is
either matched to a waiting request or joins a waiting pool at no cost.
Online matching has also been studied with
recourse~\cite{gupta2020permutation,megow2020online}, with maximization
objectives, and under other arrival
models~\cite{mehta2013online,huang2024online}.

\paragraph{Variants of matching with delays.}
Some variants change how requests arrive. Under Poisson arrivals, Mari,
Paw{\l}owski, Ren, and Sankowski proved that the greedy algorithm that
matches two requests as soon as their total waiting time exceeds their
distance is $O(1)$-competitive in expectation for the MPMD
problem~\cite{mari2023online}. Against adversarial arrivals, however, its
competitive ratio is $\Omega(m^{\log_2 1.5})$, where
$\log_2 1.5\approx 0.585$, because it reduces to the greedy heuristic for
minimum-cost perfect matching~\cite{reingold1981greedy} when all requests
arrive simultaneously.

Other variants change how many requests are matched together. Melnyk,
Wang, and Wattenhofer introduced $k$-way matching with delays, in which
requests are matched in groups of $k$, and gave a randomized algorithm on
a class of metrics called $H$-metrics~\cite{melnyk2021online}. For this
problem on $H$-metrics, Kakimura and Nakayoshi gave a deterministic
$O(mk^2)$-competitive primal-dual algorithm, where $m$ is the number of
requests~\cite{kakimura2023deterministic}. Ahmadian, Chawla, Kumar,
Purohit, and Zhang gave a deterministic $O(k\log m)$-competitive algorithm
when the cost of a group is the sum of the distances between its
requests, and a deterministic $O(k^3\log m)$-competitive algorithm for
general $H$-metrics~\cite{ahmadian2026general}.

Some variants replace the waiting time with a different delay cost. For convex delay costs $t^\alpha$ with $\alpha>1$ on uniform metric
spaces with $n$ points, the deterministic competitive ratio is
$\Theta(n)$~\cite{liu2018impatient}. For concave delay costs, Azar, Ren,
and Vainstein gave a randomized $O(\log n)$-competitive algorithm based on
HSTs for both the MPMD and the MBPMD problems~\cite{azar2021min}.
Deryckere and Umboh gave a deterministic $O(m)$-competitive primal-dual
algorithm for the MPMD problem with concave delays in general metric
spaces~\cite{deryckere2023online}. Ahmadian, Chawla, Kumar, Purohit, and
Zhang adapted their own MPMD algorithm to concave delay functions that are
known in advance and unbounded, and obtained a deterministic
$O(\log m)$-competitive algorithm~\cite{ahmadian2026general}.

Deryckere and Umboh also introduced set delays, where the delay cost at
each time is a function of the set of unmatched
requests~\cite{deryckere2023online}. For size-based delays, where this
cost depends only on the number of unmatched requests, the deterministic
competitive ratio is $\Theta(2^n)$~\cite{gan2026online}.

\paragraph{Other online problems with delays.}
The study of online problems with delays originates from the TCP
acknowledgment problem~\cite{dooly2001online}. For online service with
delays and network design with delays, randomized polylogarithmic
competitive ratios were first obtained via
HSTs~\cite{azar2017online,azar2019general}, and deterministic ones were
then obtained without tree
embeddings~\cite{azar2020beyond,touitou2023improved}.
The MBPMD problem now follows the same pattern: its known randomized
algorithms use HSTs, and Theorem~\ref{thm:main} gives a deterministic
$O(\log m)$-competitive algorithm without tree embeddings. Other problems
with delays or
deadlines include multi-level
aggregation~\cite{bienkowski2016online,buchbinder2017depth}, set cover with
delay~\cite{azar2020set}, caching with time windows~\cite{gupta2020caching},
and list update with delays~\cite{azar2024list}.

\section{Preliminaries}
\label{sec:model}

\subsection{Problem Definition}

Let $(X,d)$ be a metric space. Let $R$ and $S$ be the request set and
the server set, respectively, where $|R|=|S|=m$.
For each request or server $v\in R\cup S$, let $x_v\in X$ and
$a_v\ge0$ be its location and arrival time, respectively. We call
requests and servers \emph{items}.
When $v$ arrives, the algorithm learns its type, its location, and its
distances to all items that
arrive by time $a_v$.
The algorithm does not know future
arrivals or their locations, and it does not know the metric space or
$m$ in advance.

The algorithm must eventually output a perfect matching $M$ between $R$
and $S$. If it matches $r$ to $s$ at time $t_{rs}$, then
$t_{rs}\ge\max\{a_r,a_s\}$, and this decision cannot be changed.
The objective is to minimize the total distance of the matched pairs plus
the total waiting time of all items, that is,
\begin{equation}
  \min_{M,\,\{t_{rs}\}_{rs\in M}}\ \sum_{rs\in M}
   \bigl(d(x_r,x_s)+(t_{rs}-a_r)+(t_{rs}-a_s)\bigr).
  \label{eq:objective}
\end{equation}

\subsection{Matchings and Augmenting Paths}

A matching $M$ between $R$ and $S$ is a set of vertex-disjoint edges
between $R$ and $S$. An item is \emph{saturated} by $M$ if it is an
endpoint of an edge in $M$. Otherwise, it is \emph{free} in $M$.
An $M$-alternating path alternates between edges in $M$ and edges not in
$M$. An $M$-alternating path is $M$-augmenting if neither endpoint is
saturated by $M$. We identify a path with its edge set, and for two edge
sets $A$ and $B$, we write $A\symdiff B=(A\setminus B)\cup(B\setminus A)$
for their symmetric difference.

An $M$-augmenting path $P$ can be used to enlarge $M$. We view $P$ as a
directed path from a free request to a free server, and write $\ter(P)$
for its last vertex. Following $P$ from its free request, we update $M$
as follows:
\begin{enumerate}
\item If an edge of $P$ goes from a request to a server, it is added to
$M$.
\item If an edge of $P$ goes from a server to a request, it is removed
from $M$.
\end{enumerate}
The resulting matching is $M\symdiff P$, and its size is $|M|+1$.

\subsection{The Time-Augmented Distance}

We define the \emph{time-augmented (TA) distance} between items $u$ and
$v$ as
\begin{equation}
  D(u,v)=d(x_u,x_v)+|a_u-a_v|.
  \label{eq:ta}
\end{equation}
For an edge $e=uv$, define $D(e)=D(u,v)$, and for any edge set $F$,
define $D(F)=\sum_{e\in F}D(e)$.
The TA distance was used
in~\cite{bienkowski2017match,azar2020deterministic,kuo2025online}. The
following lemma relates it to the optimal cost.

\begin{lemma}[\cite{azar2020deterministic}]
\label{lem:offline}
If $M^*$ is a minimum-cost perfect matching between $R$ and $S$ under $D$,
then $\OPT=D(M^*)$.
\end{lemma}

\section{The VRM Algorithm}
\label{sec:algorithm}

In this section, we describe the VRM algorithm~\cite{kuo2025online} for
any metric space, in the simplified form discussed in
Section~\ref{sec:contribution}. Let $\gamma>1$ be a parameter that remains
fixed throughout the execution. We first define the augmenting paths used
by the algorithm (Section~\ref{sec:paths}) and then state the readiness
rule, which decides when a free request is matched along its augmenting
path (Section~\ref{sec:readiness}). Section~\ref{sec:events} describes
the algorithm, and
Section~\ref{sec:properties} states the basic properties used in the
analysis.

Following the notation of VRM~\cite{kuo2025online}, the algorithm
maintains an \emph{offline matching} $\Moff$ and an \emph{online matching}
$\Mout$, which is the real output matching. The offline matching is an
auxiliary matching on the items that have arrived. Both
matchings are initially empty. The algorithm can change edges in $\Moff$,
but cannot change a pair that has been added to $\Mout$.

We will see in Section~\ref{sec:events} that $\Moff$ and $\Mout$ always
saturate the same items. An item is \emph{free} at time $t$ if it has
arrived and is not saturated by $\Moff$ (equivalently, by $\Mout$) at
time $t$. Let $\Rfree(t)$ be the
set of free requests at time $t$. The set $\Rfree(t)$ does not contain
requests that have not arrived. For an arrived item $v$, its \emph{age} at time $t$ is
$\age(v,t)=t-a_v$. For a free request, the age is its current waiting time.

\subsection{Minimum Net-Cost Augmenting Paths}
\label{sec:paths}

Consider the complete bipartite graph $H$ on the requests and servers that
have arrived. For each request $r$ and server $s$, we assign the
direction and weight of their edge as follows:
\[
\begin{cases}
 r\longrightarrow s,\quad\text{weight }\gamma D(r,s),
       & rs\notin \Moff,\\
 s\longrightarrow r,\quad\text{weight }{-D(r,s)},
       & rs\in \Moff.
\end{cases}
\]
A directed path from a free request to a free server in $H$ is an
$\Moff$-augmenting path whose vertices have all arrived.

For a matching $M$ and an $M$-augmenting path
$P$, define
\begin{equation}
 \begin{split}
 D^+_M(P)&=D(P\setminus M),\qquad D^-_M(P)=D(P\cap M),\\
 \varphi_M(P)&=\gamma D^+_M(P)-D^-_M(P).
 \end{split}
 \label{eq:net}
\end{equation}
We call $\varphi_M(P)$ the \emph{$\gamma$-net-cost} of $P$.
Specifically, it is $\gamma$ times the total TA distance of the edges
added to $M$, minus the total TA distance of the edges removed from $M$.
For an $\Moff$-augmenting path $P$ from a free request to a free server,
$\varphi_{\Moff}(P)$ is also the total weight of $P$ in $H$.

For each free request $r$, let $P_r$ be an $\Moff$-augmenting path that
minimizes the $\gamma$-net-cost among all $\Moff$-augmenting paths from
$r$ to a free server. We call $P_r$ the \emph{minimum augmenting path}
of $r$, and let $\varphi_r=\varphi_{\Moff}(P_r)$ be the \emph{minimum
$\gamma$-net-cost} of $r$. Because the $\gamma$-net-cost of such a path
equals its weight in $H$, $P_r$ is a shortest path from
$r$ to the set of free servers in $H$, and $\varphi_r$ is
its length. If no free server exists, then no such path
exists, and we set $\varphi_r=+\infty$. We prove the following lemma in
Appendix~\ref{app:validity}. Its proof is similar to the corresponding
proofs for RM~\cite{raghvendra2016robust} and VRM~\cite{kuo2025online}.

\begin{lemma}
\label{lem:cycles}
The graph $H$ has no negative cycle at any state.
\end{lemma}

Thus, shortest paths exist and can be chosen simple, and $P_r$ is well
defined.

\subsection{The Readiness Rule}
\label{sec:readiness}

A free request $r$ is \emph{ready} at time $t$ if
\begin{equation}
  \gamma(t-a_r)\ge\varphi_r,
  \label{eq:ready-rule}
\end{equation}
that is, if its waiting time is at least its minimum
$\gamma$-net-cost divided by $\gamma$. When $r$ is ready, the algorithm
adds $rs$ to $\Mout$, where $s=\ter(P_r)$, and augments $\Moff$ by $P_r$. This is
the simplified readiness rule of VRM discussed in
Section~\ref{sec:contribution}~\cite[Section~4.1, Implication~1]{kuo2025online}.

Whenever $P_r$ is updated at time $t$, the algorithm checks
Eq.~\eqref{eq:ready-rule}. If $r$ is not ready, its \emph{ready time} is
\begin{equation}
  t_r^{\mathrm{rdy}}
    =t+\frac{\varphi_r-\gamma(t-a_r)}{\gamma}
    =a_r+\frac{\varphi_r}{\gamma},
  \label{eq:deadline}
\end{equation}
and $t_r^{\mathrm{rdy}}=+\infty$ if $\varphi_r=+\infty$.

\subsection{Algorithm Description}
\label{sec:events}

Following the event description of VRM~\cite[Section~4]{kuo2025online},
the execution of the algorithm is driven by three types of events:
\begin{enumerate}
\item \textbf{Event RA: A request $r$ arrives.}
Compute the minimum augmenting path $P_r$, the minimum
$\gamma$-net-cost $\varphi_r$, and the ready time of $r$.
Because $r$ is free, it cannot be an interior vertex of another free
request's augmenting path. Thus, the minimum augmenting paths and
ready times of the other free requests do not change.
\item \textbf{Event SA: A server $s$ arrives.}
Update the minimum augmenting paths, the minimum
$\gamma$-net-costs, and the ready times of all free requests.
\item \textbf{Event AU: A free request $r$ is ready.}
Let $s=\ter(P_r)$. Add $rs$ to $\Mout$ with matching time equal to the
current time, and augment $\Moff$ by $P_r$, that is, set
$\Moff\gets\Moff\symdiff P_r$. Then update the minimum augmenting
paths, the minimum $\gamma$-net-costs, and the ready times of all
remaining free requests.
\end{enumerate}
By the design of the algorithm, $\Moff$ and $\Mout$ always saturate the
same items, and every output decision is feasible.
Every free request has an edge to every free server in $H$, so
$\varphi_r<+\infty$ whenever a free server exists. After the last
arrival, there are $m-|\Moff|$ free requests and $m-|\Moff|$ free
servers, so a free server exists whenever a free request
does. Hence, after the last
arrival, every free request has a finite ready time, and the algorithm
eventually outputs a perfect matching.

Ties among simultaneous events and among minimum augmenting paths are
broken by an arbitrary but fixed rule, which makes the algorithm
deterministic, and our analysis holds for every such rule.

\subsection{Basic Properties}
\label{sec:properties}

The analysis uses the readiness rule in two ways. First, a request $r$ is
matched only when it is ready. Thus, if $r$ is matched at time $t_r$ along
the path $P_r$ with $\gamma$-net-cost $\varphi_r$, then
\begin{equation}
  \varphi_r\le\gamma(t_r-a_r).
  \label{eq:ready}
\end{equation}
Second, while a request waits, it is not ready. To make this precise, we
use the following notion.

\begin{definition}
\label{def:stable}
A time $t$ is \emph{stable} if it is not an event time, that is, no item
arrives and no augmentation is performed at time $t$.
\end{definition}

Consider a stable time $t$, a free request $r$, and an $\Moff$-augmenting
path $P$ from $r$ to a free server. Because $\varphi_r$ is the minimum
$\gamma$-net-cost of such paths, we have $\varphi_{\Moff}(P)\ge\varphi_r$.
Because $t$ is stable, Event AU does not occur at time $t$, so no free
request is ready at time $t$. In particular, $\varphi_r>\gamma(t-a_r)$ by
Eq.~\eqref{eq:ready-rule}. Therefore, we have the following lemma.

\begin{lemma}
\label{lem:stable}
At every stable time $t$, no free request is ready. In particular, every
$\Moff$-augmenting path $P$ from any free request $r$ to any free server
satisfies
\[
  \varphi_{\Moff}(P)\ge\varphi_r>\gamma(t-a_r).
\]
\end{lemma}

\section{Analysis}
\label{sec:analysis}

In this section, we prove Theorem~\ref{thm:main}. Let $t_r$ be the time
when request $r$ is matched by VRM, and let
\begin{equation}
  W_R=\sum_{r\in R}(t_r-a_r)
      =\int_0^\infty |\Rfree(t)|\,dt
  \label{eq:request-wait}
\end{equation}
be the total waiting time of requests. The second equality follows because
each free request incurs one unit of waiting cost per unit of time.
The proof has two steps. In Section~\ref{sec:ledger}, we show that
$\ALG\le(2+\frac{2\gamma}{\gamma-1})W_R$. In
Sections~\ref{sec:charging}--\ref{sec:case2}, we charge the waiting cost of every
request, at every stable time, to the edges of a fixed optimal matching
and bound the charge that each edge receives.
Section~\ref{sec:proof-main} sums these bounds to show that
$W_R\le[1+(\gamma+1)(1+\ln m)]\OPT$ and combines the two steps.

\subsection{Bounding the Total Cost by the Total Waiting Time of Requests}
\label{sec:ledger}

The following lemma relates $W_R$ to the total cost of the algorithm,
which includes the distance cost and the delay cost of both requests and
servers.

\begin{lemma}
\label{lem:ledger}
VRM satisfies
\[
  \ALG\le\left(2+\frac{2\gamma}{\gamma-1}\right)W_R.
\]
\end{lemma}
\begin{proof}
The proof is similar to the arguments
in~\cite{raghvendra2016robust}, \cite[Lemma~3.1]{kuo2025online},
and~\cite[Section~4.1, Implication~2]{kuo2025online}.
Let $P_i$ be the path used in the $i$th augmentation and let $M_i$ be the
offline matching right before this augmentation. Define
$D^+_i=D^+_{M_i}(P_i)$, $D^-_i=D^-_{M_i}(P_i)$, and
$\varphi_i=\gamma D^+_i-D^-_i$.

Because the $i$th augmentation changes the total TA distance of $\Moff$ by
$D^+_i-D^-_i$, the sum $\sum_i(D^+_i-D^-_i)$ is the total TA distance of
the final offline matching, so it is nonnegative. Because
$(\gamma-1)(D^+_i+D^-_i)+(\gamma+1)(D^+_i-D^-_i)=2\varphi_i$ for every
$i$, we have
\begin{equation}
 \begin{split}
  \sum_{i=1}^m(D^+_i+D^-_i)
    &=\frac{2}{\gamma-1}\sum_{i=1}^m\varphi_i
       -\frac{\gamma+1}{\gamma-1}\sum_{i=1}^m(D^+_i-D^-_i)\\
    &\le\frac{2}{\gamma-1}\sum_{i=1}^m\varphi_i
    \le\frac{2\gamma}{\gamma-1}\,W_R.
 \end{split}
  \label{eq:ledger}
\end{equation}
The last inequality follows from Eq.~\eqref{eq:ready}, because
$\varphi_i\le\gamma(t_r-a_r)$, where $r$ is the starting request of $P_i$.

If the $i$th augmentation outputs the pair $rs$ at time $t_r$, then
$t_r\ge\max\{a_r,a_s\}$, and the triangle inequality gives
$D(r,s)\le D(P_i)=D^+_i+D^-_i$. The distance and delay cost of this pair is
\begin{align*}
 d(x_r,x_s)&+(t_r-a_r)+(t_r-a_s)\\
   &=D(r,s)+2\bigl(t_r-\max\{a_r,a_s\}\bigr)\\
   &\le D(r,s)+2(t_r-a_r)\le D^+_i+D^-_i+2(t_r-a_r).
\end{align*}
Summing over all $m$ pairs and applying Eq.~\eqref{eq:ledger}, we have
\[
  \ALG\le\frac{2\gamma}{\gamma-1}W_R+2W_R
       =\left(2+\frac{2\gamma}{\gamma-1}\right)W_R,
\]
as desired.
\end{proof}

\subsection{Optimal Paths and the Charging Scheme}
\label{sec:charging}

To bound $W_R$, we compare the offline matching with a fixed optimal
matching under the TA distance. Let $M^*$ be a minimum-cost perfect matching
between $R$ and $S$ under $D$, so that $D(M^*)=\OPT$ by
Lemma~\ref{lem:offline}. Let $M_t$ be the offline matching at time $t$.
Consider a free request $r$ at time $t$. Because $r$ is not saturated by
$M_t$ and $M^*$ is perfect, the component of $M_t\symdiff M^*$ containing
$r$ is an $M_t$-augmenting path from $r$ to a server. We call this path the
\emph{optimal path} of $r$ at time $t$, denote it by $P^*_r(t)$, and call
its last vertex the \emph{terminal server}. The terminal server is not
saturated by $M_t$, so it is either free or not yet arrived.

\begin{lemma}
\label{lem:structure}
At any time $t$, the optimal paths of different free requests share no
edge, and every interior vertex of an optimal path has arrived.
\end{lemma}
\begin{proof}
The optimal paths of different free requests are different components of
$M_t\symdiff M^*$, so they share no edge. Next, an optimal path alternates between edges of $M^*$ and edges of $M_t$, so
every interior vertex is incident to an edge of $M_t$. Because $M_t$ is a
matching on the items that have arrived, every interior vertex has
arrived.
\end{proof}

By Lemma~\ref{lem:structure}, the only vertex of $P^*_r(t)$ that may not
have arrived is its terminal server.

\paragraph{The charging scheme.}
There are at most $2m$ arrival times and at most $m$
augmentations, so there are only finitely many event times. Thus, the
waiting cost incurred at event times is zero, and it suffices to charge
the waiting cost at stable times. We consider a stable time $t$ and a free
request $r$. The waiting cost of $r$
increases at rate one, and we charge it as follows.
\begin{itemize}
\item \textbf{Case 1: the terminal server of $P^*_r(t)$ has not arrived.}
We charge the waiting cost of $r$ to the last edge of $P^*_r(t)$.
Because the terminal server is not saturated by $M_t$, this edge is
not in $M_t$. It lies in $M_t\symdiff M^*$, so it is an edge of
$M^*$. By
Lemma~\ref{lem:structure}, at any time, each edge is charged by at most
one request. Let $W_1$ be the total waiting cost charged in this case over
all requests and over time. Section~\ref{sec:case1} shows that
$W_1\le\OPT$.
\item \textbf{Case 2: the terminal server of $P^*_r(t)$ has arrived.}
Let $D^*=D(P^*_r(t)\cap M^*)$ be the total TA distance of the edges of $M^*$
on $P^*_r(t)$. For each edge $e\in P^*_r(t)\cap M^*$, we charge the waiting
cost of $r$ to $e$ at rate
\begin{equation}
  \frac{D(e)}{D^*}.
  \label{eq:charge-rule}
\end{equation}
Lemma~\ref{lem:window} in Section~\ref{sec:case2} gives $D^*\ge\age(r,t)$,
and $\age(r,t)>0$ because no item arrives at the stable time $t$. Thus,
the rate is well defined. Figure~\ref{fig:charge-rate}(a) illustrates this rule. By
Lemma~\ref{lem:structure}, at any time, each edge is charged by at most
one request. For each edge
$e\in M^*$, let $q_e(t)$ be the charge rate assigned to $e$ in this
case at a stable time $t$, with $q_e(t)=0$ at event times.
\end{itemize}
Every unit of $W_R$ is charged
by exactly one of the above rules. Therefore,
\begin{equation}
  W_R=W_1+\sum_{e\in M^*}\int_0^\infty q_e(t)\,dt.
  \label{eq:decomposition}
\end{equation}
It remains to bound $W_1$ and
$\int_0^\infty q_e(t)\,dt$ for every $e\in M^*$. We treat these bounds in
Sections~\ref{sec:case1} and~\ref{sec:case2}.

\subsection{Case 1: The Terminal Server Has Not Arrived}
\label{sec:case1}

\begin{lemma}
\label{lem:future}
The total waiting cost charged in Case~1 satisfies $W_1\le\OPT$.
\end{lemma}
\begin{proof}
In Case~1, a free request charges its waiting cost to the last
edge of its optimal path. Fix an edge $e=ps\in M^*$ that is charged in
Case~1, where $p\in R$ and $s\in S$. If a free request $r$ charges $e$ in
Case~1 at time $t$, then $s$ is the terminal server of $P^*_r(t)$, so $s$
has not arrived at time $t$. By Lemma~\ref{lem:structure}, $p$ has
arrived, since it is $r$ or an interior vertex of $P^*_r(t)$. Thus,
$a_p\le t<a_s$, so $p$ arrives before $s$, and $e$ is charged only during
$[a_p,a_s]$.

By Lemma~\ref{lem:structure}, $e$ is charged by at most one request at
any stable time. Because each free request incurs waiting cost at unit
rate, the total charge to $e$ in Case~1 is at most $a_s-a_p\le D(p,s)$.
An edge of $M^*$ that is never charged in Case~1 contributes nothing, so
summing over all edges of $M^*$ gives
$W_1\le\sum_{ps\in M^*}D(p,s)=\OPT$.
\end{proof}

\subsection{Case 2: The Terminal Server Has Arrived}
\label{sec:case2}

In Case~2, an edge $e\in P^*_r(t)\cap M^*$ is charged at rate $D(e)/D^*$.
This rate is large only when $D^*$ is small, so bounding the total charge
to $e$ amounts to showing that $D^*$ cannot be small. The next lemma gives
two lower bounds on $D^*$.

\begin{lemma}
\label{lem:window}
Fix a stable time $t$ and a free request $r$ such that the terminal server
of $P^*=P^*_r(t)$ has arrived, and let $D^*=D(P^*\cap M^*)$. Then
\[
  D^*\ge\age(r,t)
  \qquad\text{and}\qquad
  D^*\ge\frac{\age(v,t)}{\gamma+1}\ \text{ for every vertex $v$ of $P^*$.}
\]
\end{lemma}
\begin{proof}
Let $D^-=D(P^*\cap M_t)$, so that $D(P^*)=D^*+D^-$. By
Lemma~\ref{lem:structure} and the assumption, all vertices of $P^*$ have
arrived. The terminal server of $P^*$ is not saturated by $M_t$, so it is
free. Thus, $P^*$ is an $M_t$-augmenting path from the free request $r$ to a
free server, and its $\gamma$-net-cost is
$\varphi_{M_t}(P^*)=\gamma D^*-D^-$. We prove the two bounds as follows.
\begin{enumerate}
\item By Lemma~\ref{lem:stable}, we have $\gamma D^*-D^->\gamma\age(r,t)$. Because
$D^-\ge0$, this gives the first bound $D^*\ge\age(r,t)$.
\item Rearranging the inequality $\gamma D^*-D^->\gamma\age(r,t)$ above gives
$D^-<\gamma(D^*-\age(r,t))$. Thus,
$D(P^*)=D^*+D^-<(\gamma+1)D^*-\gamma\age(r,t)$.
\item Let $v$ be a vertex of $P^*$, and let $P^*[r,v]$ be the subpath of $P^*$ from $r$ to $v$. By the
definition of $D$ and the triangle inequality for $D$, we have
$|a_r-a_v|\le D(r,v)\le D(P^*[r,v])\le D(P^*)$. By the bound on $D(P^*)$
above, $\gamma>1$, and $\age(r,t)\ge0$, we have
\[
  \age(v,t)=\age(r,t)+(a_r-a_v)\le\age(r,t)+D(P^*)
  <(\gamma+1)D^*-(\gamma-1)\age(r,t)\le(\gamma+1)D^*,
\]
which is the second bound.\qedhere
\end{enumerate}
\end{proof}

We now fix an edge $e\in M^*$ and bound the total charge that it
receives.

\begin{lemma}
\label{lem:edge}
For every edge $e\in M^*$, the total charge received by $e$ in Case~2
satisfies
\[
  \int_0^\infty q_e(t)\,dt\le(\gamma+1)\Bigl(\frac{\OPT}{m}+D(e)\ln m\Bigr).
\]
\end{lemma}
\begin{proof}
Choose one endpoint $v$ of $e$. Let $b_e=a_v$
be the arrival time of this endpoint. We use the same endpoint $v$
throughout the analysis of $e$.
Consider a stable time $t$ with $q_e(t)>0$. By Lemma~\ref{lem:structure},
exactly one request, which we denote by $r_e(t)$, charges $e$ at time $t$. Let $P^*_e(t)=P^*_{r_e(t)}(t)$ be its
optimal path, and let $D^*_e(t)=D(P^*_e(t)\cap M^*)$. By
Eq.~\eqref{eq:charge-rule}, we have $q_e(t)=D(e)/D^*_e(t)$. Both $r_e(t)$
and $P^*_e(t)$ may change as $t$ varies.

The waiting cost of $r_e(t)$ increases at rate one and is split among the
edges of $P^*_e(t)\cap M^*$. Because $r_e(t)$ is the only request that
charges $e$ at time $t$, we have $q_e(t)\le1$.
Moreover, $v$ is a vertex of $P^*_e(t)$, and in Case~2 every vertex of
$P^*_e(t)$ has arrived. Thus, $v$ has arrived. Because no item arrives at
the stable time $t$, we have $\age(v,t)=t-b_e>0$. The second bound of
Lemma~\ref{lem:window}, applied to $r_e(t)$, gives
\begin{equation}
  D^*_e(t)\ge(t-b_e)/(\gamma+1).
  \label{eq:edge-lower}
\end{equation}

Next, we bound the time interval during which $e$ is charged. Because
$D^*_e(t)\le D(M^*)=\OPT$, Eq.~\eqref{eq:edge-lower} gives
$t\le b_e+(\gamma+1)\OPT$. Together with $t>b_e$, this shows that $e$
is charged only during $(b_e,\,b_e+(\gamma+1)\OPT]$. Combining this with
$q_e(t)\le1$, Eq.~\eqref{eq:edge-lower}, and $q_e(t)=D(e)/D^*_e(t)$, we
have
\begin{equation}
  \begin{aligned}
    q_e(t)&\le\min\Bigl\{1,\,\frac{(\gamma+1)D(e)}{t-b_e}\Bigr\}
      &&\text{if } t\in(b_e,\,b_e+(\gamma+1)\OPT],\\
    q_e(t)&=0 &&\text{otherwise}.
  \end{aligned}
  \label{eq:envelope}
\end{equation}

Different requests may charge $e$ during different parts of this interval,
and $e$ need not receive charge throughout the interval. The upper bound
in Eq.~\eqref{eq:envelope} holds at every time, regardless of which
request is charging $e$. Figure~\ref{fig:charge-rate}(b) illustrates
$q_e(t)$ and this bound.

Integrating Eq.~\eqref{eq:envelope} bounds the total charge that $e$
receives from all requests. We split the integral at
$b_e+(\gamma+1)\,{\OPT}/m$. We bound the minimum in
Eq.~\eqref{eq:envelope} by $1$ before this time and by
$(\gamma+1)D(e)/(t-b_e)$ after it, and obtain
\begin{align}
 \int_0^\infty q_e(t)\,dt
 &\le \int_{b_e}^{b_e+(\gamma+1)\,{\OPT}/m}1\,dt
     +\int_{b_e+(\gamma+1)\,{\OPT}/m}^{b_e+(\gamma+1)\OPT}
       \frac{(\gamma+1)D(e)}{t-b_e}\,dt \notag\\
 &=(\gamma+1)\Bigl(\frac{\OPT}{m}+D(e)\ln m\Bigr).
 \label{eq:integral}
\end{align}
\end{proof}

\begin{widefigure}[tbp]
  \centering
  \begingroup
  \definecolor{vrmBlue}{HTML}{0072B2}
  \definecolor{vrmOrange}{HTML}{D55E00}
  \definecolor{vrmGreen}{HTML}{009E73}
  \definecolor{vrmPurple}{HTML}{CC79A7}
  \definecolor{vrmSky}{HTML}{56B4E9}
  \begin{tikzpicture}[
    x=1cm,y=1cm,
    font=\fontsize{9}{11}\selectfont,
    request/.style={circle,draw=black,fill=white,line width=.6pt,
                    minimum size=5.5mm,inner sep=0pt},
    server/.style={rectangle,draw=black,fill=white,line width=.6pt,
                   minimum size=5.5mm,inner sep=0pt},
    optedge/.style={draw=vrmBlue,line width=1.1pt},
    offedge/.style={draw=black!70,line width=.8pt,densely dashed},
    panel/.style={anchor=west,font=\fontsize{9}{11}\selectfont\bfseries},
    axis/.style={draw=black!75,line width=.6pt,
                 -{Stealth[length=1.8mm,width=1.3mm]}},
    qstep/.style={line width=1.4pt},
    bound/.style={draw=black!60,line width=.9pt,dashed},
  ]
    \useasboundingbox (0,-.45) rectangle (14,8.45);
    \node[panel] at (0,8.2)
      {(a) At a stable time $t$: the optimal path $P^*_e(t)$ of the request $r_e(t)$};
    \node[request,fill=vrmGreen!20] (r)  at (0.7,6.8) {};
    \node[server]  (s1) at (3.3,6.8) {};
    \node[request] (p1) at (5.2,6.8) {};
    \node[server]  (s2) at (8.0,6.8) {};
    \node[request] (p2) at (9.9,6.8) {};
    \node[server]  (s)  at (13.3,6.8) {};
    \draw[optedge] (r) -- node[above=2pt,text=black] {$e_1$} (s1);
    \draw[offedge] (s1) -- (p1);
    \draw[optedge,line width=2.6pt] (p1) -- node[above=2pt,text=black] {$e$} (s2);
    \draw[offedge] (s2) -- (p2);
    \draw[optedge] (p2) -- node[above=2pt,text=black] {$e_2$} (s);
    \node[below=3pt,text=vrmBlue] at ($(r)!.5!(s1)$) {$D(e_1)/D^*_e(t)$};
    \node[below=3pt,text=vrmBlue] at ($(p1)!.5!(s2)$) {$D(e)/D^*_e(t)$};
    \node[below=3pt,text=vrmBlue] at ($(p2)!.5!(s)$) {$D(e_2)/D^*_e(t)$};
    \node[above=3pt] at (r.north) {$r_e(t)$};
    \node[above=3pt] at (s.north) {free server};
    \node[anchor=west,text=black!70] at (0.1,5.9)
      {Share of the unit waiting rate of $r_e(t)$ received by each $M^*$-edge,
       where $D^*_e(t)=D(e_1)+D(e)+D(e_2)$.};
    \node[panel] at (0,5.2) {(b) The charge rate $q_e(t)$ of the edge $e$ over time};
    \draw[bound] (0.8,4.0) -- (2.6,4.0)
      -- plot[domain=2.6:13.4,samples=120,smooth] (\x,{0.4+6.48/(\x-0.8)});
    \node[anchor=south,text=black!65] at (1.7,4.02) {$1$};
    \node[anchor=south west,text=black!65] at (4.3,2.3)
      {$(\gamma+1)D(e)/(t-b_e)$};
    \fill[vrmOrange!18] (1.1,0.4) rectangle (1.22,3.25);
    \draw[qstep,draw=vrmOrange] (1.1,3.25) -- (1.22,3.25);
    \fill[vrmOrange!18] (1.22,0.4) rectangle (1.6,2.25);
    \draw[qstep,draw=vrmOrange] (1.22,2.25) -- (1.6,2.25);
    \fill[vrmOrange!18] (1.6,0.4) rectangle (2.3,1.5);
    \draw[qstep,draw=vrmOrange] (1.6,1.5) -- (2.3,1.5);
    \draw[draw=vrmOrange,line width=.6pt] (1.22,2.25) -- (1.22,3.25);
    \draw[draw=vrmOrange,line width=.6pt] (1.6,1.5) -- (1.6,2.25);
    \node[text=vrmOrange!85!black] at (1.95,0.95) {$r_1$};
    \fill[vrmPurple!18] (3.5,0.4) rectangle (3.62,2.45);
    \draw[qstep,draw=vrmPurple] (3.5,2.45) -- (3.62,2.45);
    \fill[vrmPurple!18] (3.62,0.4) rectangle (3.9,2.05);
    \draw[qstep,draw=vrmPurple] (3.62,2.05) -- (3.9,2.05);
    \fill[vrmPurple!18] (3.9,0.4) rectangle (5.2,1.15);
    \draw[qstep,draw=vrmPurple] (3.9,1.15) -- (5.2,1.15);
    \draw[draw=vrmPurple,line width=.6pt] (3.62,2.05) -- (3.62,2.45);
    \draw[draw=vrmPurple,line width=.6pt] (3.9,1.15) -- (3.9,2.05);
    \node[text=vrmPurple!85!black] at (4.55,0.775) {$r_2$};
    \fill[vrmGreen!20] (5.6,0.4) rectangle (6.2,1.02);
    \draw[qstep,draw=vrmGreen] (5.6,1.02) -- (6.2,1.02);
    \fill[vrmGreen!20] (6.2,0.4) rectangle (6.8,1.12);
    \draw[qstep,draw=vrmGreen] (6.2,1.12) -- (6.8,1.12);
    \fill[vrmGreen!20] (6.8,0.4) rectangle (7.8,0.96);
    \draw[qstep,draw=vrmGreen] (6.8,0.96) -- (7.8,0.96);
    \draw[draw=vrmGreen,line width=.6pt] (6.2,1.02) -- (6.2,1.12);
    \draw[draw=vrmGreen,line width=.6pt] (6.8,0.96) -- (6.8,1.12);
    \node[text=vrmGreen!75!black] at (5.9,0.71) {$r_3$};
    \fill[vrmSky!25] (8.3,0.4) rectangle (8.8,1.03);
    \draw[qstep,draw=vrmSky] (8.3,1.03) -- (8.8,1.03);
    \fill[vrmSky!25] (8.8,0.4) rectangle (9.45,0.94);
    \draw[qstep,draw=vrmSky] (8.8,0.94) -- (9.45,0.94);
    \draw[draw=vrmSky,line width=.6pt] (8.8,0.94) -- (8.8,1.03);
    \node[text=vrmSky!70!black] at (9.125,0.67) {$r_4$};
    \draw[axis] (0.8,0.4) -- (13.7,0.4) node[right] {$t$};
    \draw[axis] (0.8,0.4) -- (0.8,4.5) node[above] {rate};
    \foreach \xx in {0.8,3.32,13.4}
      \draw[draw=black!70,line width=.6pt] (\xx,0.35) -- (\xx,0.45);
    \draw[draw=black!35,densely dotted,line width=.5pt] (3.32,0.4) -- (3.32,2.971);
    \draw[draw=black!35,densely dotted,line width=.5pt] (13.4,0.4) -- (13.4,0.914);
    \node[anchor=north] at (0.8,0.3) {$b_e$};
    \node[anchor=north] at (3.32,0.3) {$b_e+(\gamma+1)\,{\OPT}/m$};
    \node[anchor=north] at (13.4,0.3) {$b_e+(\gamma+1)\OPT$};
    \draw[qstep,draw=black!70] (8.6,4.2) -- (9.2,4.2);
    \node[anchor=west] at (9.25,4.2) {$q_e(t)$, colored by the request $r_e(t)$};
    \draw[bound] (8.6,3.75) -- (9.2,3.75);
    \node[anchor=west] at (9.25,3.75) {upper bound in Eq.~\eqref{eq:envelope}};
  \end{tikzpicture}
  \endgroup
  \caption{Charges received by a fixed edge $e\in M^*$ in Case~2. In (a),
  solid edges belong to $M^*$, and dashed edges belong to $M_t$. In (b),
  the charge rate of each request changes when an augmentation changes its
  optimal path, and $b_e+(\gamma+1)\,{\OPT}/m$ is the time where the proof of
  Lemma~\ref{lem:edge} splits the integral.}
  \Description{Panel (a) shows an alternating path from a free request to
  a free server with three optimal edges, each labeled with its share of
  the waiting rate. Panel (b) shows a step function below a curve that is
  constant and then decreases like the reciprocal of the elapsed time.}
  \label{fig:charge-rate}
\end{widefigure}

\subsection{Proof of Theorem~\ref{thm:main}}
\label{sec:proof-main}

By Eq.~\eqref{eq:decomposition} and Lemmas~\ref{lem:future}
and~\ref{lem:edge}, and because $M^*$ has $m$ edges and
$\sum_{e\in M^*}D(e)=\OPT$ by Lemma~\ref{lem:offline}, the total waiting
time of requests satisfies
\[
  W_R
   \le\OPT+\sum_{e\in M^*}(\gamma+1)\Bigl(\frac{\OPT}{m}+D(e)\ln m\Bigr)
   =\bigl[1+(\gamma+1)(1+\ln m)\bigr]\OPT.
\]

Combining this bound with Lemma~\ref{lem:ledger}, we have
\[
  \ALG
    \le\left(2+\frac{2\gamma}{\gamma-1}\right)W_R
    \le\left(2+\frac{2\gamma}{\gamma-1}\right)
       \bigl[1+(\gamma+1)(1+\ln m)\bigr]\OPT.
\]
For $\gamma=2$, the coefficient
is $6(4+3\ln m)=24+18\ln m$. Finally, the algorithm does not need to know
the metric space or $m$ in advance, since $M^*$, $\OPT$, and $m$ are used only in the analysis.

\section{Extension to Non-Bipartite Matching with Delays}
\label{sec:nonbipartite}

In this section, we extend VRM and its analysis to the MPMD problem, in
which any two requests can be matched. Let $R$ be the set of requests,
which arrive over time in a metric space, and let $|R|=2m$, so that $m$
is the number of pairs in a perfect matching. We use the same notation
for their locations $x_v$, arrival times $a_v$, and TA distances
$D(u,v)=d(x_u,x_v)+|a_u-a_v|$. The objective is to minimize the total
distance of the matched pairs plus the total waiting time of all
requests.
Lemma~\ref{lem:offline} also holds for the MPMD
problem~\cite{azar2020deterministic}. That is, $D(M^*)=\OPT$ for every
minimum-cost perfect matching $M^*$ on $R$ under $D$.

We call our extension \emph{non-bipartite VRM}. Because there are no
servers, both endpoints of an augmenting path are requests, and each of
them can initiate an augmentation. The analysis uses the same optimal
paths and charging scheme as in Section~\ref{sec:analysis}. The main
difference is that we now charge the waiting time of both endpoints of
an optimal path. Thus, two requests can charge the same edge of $M^*$ at
the same time. We obtain the following result.

\begin{theorem}
\label{thm:nonbipartite}
For every fixed $\gamma>1$, the non-bipartite VRM algorithm described
below is a deterministic online algorithm for the MPMD problem in any
metric space. On every input, its cost satisfies
\[
  \ALG\le\frac{2\gamma}{\gamma-1}
  \bigl[1+2(\gamma+1)(1+\ln m)\bigr]\OPT.
\]
In particular, for $\gamma=2$, it is
$(28+24\ln m)$-competitive. The algorithm does not need to know the
metric space or $m$ in advance.
\end{theorem}

\subsection{The Algorithm}
\label{sec:nb-algorithm}

As in Section~\ref{sec:algorithm}, the algorithm maintains an offline
matching $\Moff$ and an online matching $\Mout$, both initially empty.
A request is free at time $t$ if it has arrived and is not saturated by
$\Moff$ (equivalently, by $\Mout$) at time $t$.
Because any two requests can be matched, $\Moff$ can contain an edge
between any two distinct arrived requests. An $\Moff$-augmenting path is a
simple alternating path between two distinct requests that are free in
$\Moff$. For a matching $M$ and an edge set $F$, define
\[
  \varphi_M(F)=\gamma D(F\setminus M)-D(F\cap M).
\]
We call $\varphi_M(F)$ the \emph{$\gamma$-net-cost} of $F$. For an
$M$-augmenting path, this agrees with Eq.~\eqref{eq:net}.

For each free request $u$, let $P_u$ be an $\Moff$-augmenting path that
minimizes the $\gamma$-net-cost among all $\Moff$-augmenting paths from
$u$ to another free request. We call $P_u$ the \emph{minimum augmenting
path} of $u$, and let $\varphi_u=\varphi_{\Moff}(P_u)$ be the
\emph{minimum $\gamma$-net-cost} of $u$. If there is no other free request,
we set $\varphi_u=+\infty$. Otherwise, the edge from $u$ to another free
request is a candidate, and there are only finitely many simple paths on
the arrived requests. Thus, $P_u$ is well defined.

A free request $u$ is \emph{ready} at time $t$ if
\begin{equation}
  \gamma(t-a_u)\ge\varphi_u.
  \label{eq:nb-ready}
\end{equation}
Whenever $P_u$ is updated, the algorithm checks
Eq.~\eqref{eq:nb-ready}. If $u$ is not ready, its \emph{ready time} is
$a_u+\varphi_u/\gamma$, which is $+\infty$ if $\varphi_u=+\infty$.

The execution of the algorithm is driven by two types of events:
\begin{enumerate}
\item \textbf{Event RA: A request $v$ arrives.}
Update the minimum augmenting paths, the minimum $\gamma$-net-costs, and
the ready times of all free requests, including $v$. Because $v$ can be
the other endpoint of the minimum augmenting paths of the other free
requests, these requests must also be updated, unlike in Event RA of
Section~\ref{sec:events}.
\item \textbf{Event AU: A free request $u$ is ready.}
Let $v$ be the other endpoint of $P_u$. Add $uv$ to $\Mout$ with matching
time equal to the current time, and set $\Moff\gets\Moff\symdiff P_u$.
Then update the minimum augmenting paths, the minimum
$\gamma$-net-costs, and the ready times of all remaining free requests.
\end{enumerate}
Ties among simultaneous events and among the choices in Steps~1
and~2 of Section~\ref{sec:nb-implementation} are broken by an arbitrary
but fixed rule, which makes the algorithm deterministic, and our analysis
holds for every such rule. The matchings $\Moff$ and $\Mout$ always
saturate the same requests, so every output decision is feasible, and the
algorithm eventually outputs a perfect matching.

As in Definition~\ref{def:stable}, a time is \emph{stable} if no request
arrives and no augmentation is performed at that time. For an arrived
request $v$, let $\age(v,t)=t-a_v$ be its age at time $t$. Consider a stable
time $t$ and an $\Moff$-augmenting path $P$ between free requests $u$ and
$v$. Because $P$ can be traversed in either direction, it is a candidate
for the minimum augmenting paths of both $u$ and $v$. Because $t$ is
stable, Event AU does not occur at time $t$, so no free request is ready at
time $t$. Thus, by Eq.~\eqref{eq:nb-ready}, we have
\begin{equation}
  \varphi_{\Moff}(P)\ge\varphi_u>\gamma\age(u,t),
  \qquad
  \varphi_{\Moff}(P)\ge\varphi_v>\gamma\age(v,t).
  \label{eq:nb-stable}
\end{equation}

\subsection{Computing Minimum Augmenting Paths}
\label{sec:nb-implementation}

The graph $H$ of Section~\ref{sec:paths} relies on a fixed bipartition.
We instead obtain the minimum augmenting paths from minimum-weight
perfect matchings in general graphs. Consider a state with offline
matching $M$, and let $U$ be the set of requests saturated by $M$. For
two distinct free requests $u$ and $v$, let $G_{uv}$ be the complete
graph on $U\cup\{u,v\}$ with edge weights
\[
  w_M(e)=
  \begin{cases}
    D(e),&e\in M,\\
    \gamma D(e),&e\notin M.
  \end{cases}
\]
For every perfect matching $N$ in $G_{uv}$, the edge set $N\symdiff M$
consists of one $M$-augmenting path from $u$ to $v$ and vertex-disjoint
simple $M$-alternating even cycles.

To update the minimum augmenting paths and the minimum
$\gamma$-net-costs in Events RA and AU, the algorithm performs the
following two steps.
\begin{enumerate}
\item For every pair of distinct free requests $u$ and $v$, compute a
minimum-weight perfect matching $N_{uv}$ in $G_{uv}$, and let $Q_{uv}$
be the $M$-augmenting path from $u$ to $v$ in $N_{uv}\symdiff M$.
\item For every free request $u$, choose a free request $v\ne u$ that
minimizes $\varphi_M(Q_{uv})$, and set $P_u=Q_{uv}$ and
$\varphi_u=\varphi_M(Q_{uv})$. If $u$ is the only free request, set
$\varphi_u=+\infty$.
\end{enumerate}

We now show that the two steps compute the minimum augmenting path $P_u$
and the minimum $\gamma$-net-cost $\varphi_u$ of every free request $u$.
Because Step~2 minimizes $\varphi_M(Q_{uv})$ over $v$, it suffices to
show that $Q_{uv}$ minimizes the $\gamma$-net-cost among all
$M$-augmenting paths from $u$ to $v$. We first relate the weights in
$G_{uv}$ to $\gamma$-net-costs.

\begin{lemma}
\label{lem:nb-weight}
Let $M$ be the offline matching at some state, and let $u$ and $v$ be
distinct free requests. Every perfect matching $N$ in $G_{uv}$ satisfies
$w_M(N)-w_M(M)=\varphi_M(N\symdiff M)$.
\end{lemma}
\begin{proof}
We have
\[
  w_M(N)-w_M(M)=\gamma D(N\setminus M)+D(N\cap M)-D(M)
   =\gamma D(N\setminus M)-D(M\setminus N)=\varphi_M(N\symdiff M).
\]
\end{proof}

For every $M$-augmenting path $P$ from $u$ to $v$, the matching
$M\symdiff P$ is a perfect matching in $G_{uv}$, and
Lemma~\ref{lem:nb-weight} gives $w_M(M\symdiff P)=w_M(M)+\varphi_M(P)$.
Thus, to show that $Q_{uv}$ minimizes the $\gamma$-net-cost among all
$M$-augmenting paths from $u$ to $v$, it suffices to show that
$M\symdiff Q_{uv}$ is a minimum-weight perfect matching in $G_{uv}$.

The matching $N_{uv}$ is a minimum-weight perfect matching in $G_{uv}$.
It differs from $M\symdiff Q_{uv}$ only on the alternating cycles of
$N_{uv}\symdiff M$. Lemma~\ref{lem:nb-optimal}
shows that if no simple $M$-alternating even cycle has negative
$\gamma$-net-cost, these cycles cannot make $N_{uv}$ lighter than
$M\symdiff Q_{uv}$, so $M\symdiff Q_{uv}$ is also a minimum-weight
perfect matching.
Lemma~\ref{lem:nb-cycles} shows by induction on the events that the
offline matching satisfies this condition at every state.

\begin{lemma}
\label{lem:nb-optimal}
Let $M$ be the offline matching at some state, and suppose that every
simple $M$-alternating even cycle has nonnegative $\gamma$-net-cost.
Then for every pair of distinct free requests $u$ and $v$, the matching
$M\symdiff Q_{uv}$ is a minimum-weight perfect matching in $G_{uv}$.
\end{lemma}
\begin{proof}
The edge set $N_{uv}\symdiff M$ consists of $Q_{uv}$ and simple
$M$-alternating even cycles, and each of these cycles has nonnegative
$\gamma$-net-cost by assumption. Because $\varphi_M$ is additive over
disjoint edge sets, Lemma~\ref{lem:nb-weight} gives
\[
  w_M(M\symdiff Q_{uv})-w_M(M)=\varphi_M(Q_{uv})
   \le\varphi_M(N_{uv}\symdiff M)=w_M(N_{uv})-w_M(M).
\]
Because $M\symdiff Q_{uv}$ is a perfect matching in $G_{uv}$, it is a
minimum-weight perfect matching in $G_{uv}$.
\end{proof}

The induction step of Lemma~\ref{lem:nb-cycles} compares
$\gamma$-net-costs with respect to the offline matchings before and
after an augmentation, and we make this comparison edge by edge.
For a matching $L$ and an edge $e$, let $c_L(e)=-D(e)$ if $e\in L$ and
$c_L(e)=\gamma D(e)$ otherwise. Then $c_L(e)$ is the $\gamma$-net-cost of
switching $e$ with respect to $L$, and $\varphi_L(F)$ is the sum of
$c_L(e)$ over $e\in F$.

\begin{lemma}
\label{lem:nb-edge-costs}
Let $M$ and $M'$ be matchings, and let $Q=M\symdiff M'$.
\begin{enumerate}
\item[(a)] Every edge $e\notin Q$ satisfies $c_{M'}(e)=c_M(e)$.
\item[(b)] Every edge $e\in Q$ satisfies
$c_M(e)+c_{M'}(e)=(\gamma-1)D(e)$.
\end{enumerate}
\end{lemma}
\begin{proof}
(a) Because $Q=M\symdiff M'$, an edge outside $Q$ belongs to $M$ if and
only if it belongs to $M'$.

(b) An edge $e\in Q$ belongs to exactly one of $M$ and $M'$. Thus, one of
$c_M(e)$ and $c_{M'}(e)$ is $-D(e)$, and the other is $\gamma D(e)$.
\end{proof}

\begin{lemma}
\label{lem:nb-cycles}
At every state of non-bipartite VRM, every simple $\Moff$-alternating
even cycle $C$ satisfies $\varphi_{\Moff}(C)\ge0$.
\end{lemma}
\begin{proof}
We use induction on the events. Initially, $\Moff$ is empty, so there is
no alternating cycle. Every vertex of an alternating cycle is saturated,
and a newly arriving request is free. Thus, an arrival preserves the
statement.

Consider an augmentation for a free request $u$, and let $v$ be the free
request chosen for $u$ in Step~2, so that the augmentation is along
$Q_{uv}$. Let $M$ be the offline matching before the augmentation, and
let $M'=M\symdiff Q_{uv}$. By the induction hypothesis and
Lemma~\ref{lem:nb-optimal}, $M'$ is a minimum-weight perfect matching in
$G_{uv}$.

Suppose, for contradiction, that some simple $M'$-alternating even cycle
$C$ satisfies $\varphi_{M'}(C)<0$. Because $C$ is $M'$-alternating,
$N=M'\symdiff C$ is a matching that saturates the same requests as $M'$,
so $N$ is also a perfect matching in $G_{uv}$. Because $M'$ is a
minimum-weight perfect matching in $G_{uv}$, it suffices to show that
$w_M(N)<w_M(M')$ to reach a contradiction.

We have
$N\symdiff M=M'\symdiff C\symdiff M=M\symdiff Q_{uv}\symdiff C\symdiff M
=Q_{uv}\symdiff C$ and $M'\symdiff M=Q_{uv}$. Applying
Lemma~\ref{lem:nb-weight} to $N$ and to $M'$, we obtain
\begin{align*}
  w_M(N)-w_M(M')
   &=\bigl(w_M(N)-w_M(M)\bigr)-\bigl(w_M(M')-w_M(M)\bigr)\\
   &=\varphi_M(Q_{uv}\symdiff C)-\varphi_M(Q_{uv})\\
   &=\sum_{e\in C\setminus Q_{uv}}c_M(e)-\sum_{e\in Q_{uv}\cap C}c_M(e).
\end{align*}
Because $Q_{uv}=M\symdiff M'$,
Lemma~\ref{lem:nb-edge-costs}(a) gives $c_M(e)=c_{M'}(e)$ for every
$e\in C\setminus Q_{uv}$, and Lemma~\ref{lem:nb-edge-costs}(b) gives
$-c_M(e)=c_{M'}(e)-(\gamma-1)D(e)$ for every $e\in Q_{uv}\cap C$.
Therefore,
\begin{align*}
  w_M(N)-w_M(M')
   &=\sum_{e\in C\setminus Q_{uv}}c_{M'}(e)
     +\sum_{e\in Q_{uv}\cap C}\bigl(c_{M'}(e)-(\gamma-1)D(e)\bigr)\\
   &=\varphi_{M'}(C)-(\gamma-1)D(Q_{uv}\cap C)<0,
\end{align*}
where the inequality uses $\varphi_{M'}(C)<0$ and $\gamma>1$.
\end{proof}

\subsection{Bounding the Total Cost by the Total Waiting Time}
\label{sec:nb-ledger}

Let $t_r$ be the time when request $r$ is matched, and let $\Rfree(t)$
be the set of free requests at time $t$. The total waiting
time of all requests is
\[
  W_R=\sum_{r\in R}(t_r-a_r)=\int_0^\infty |\Rfree(t)|\,dt,
\]
where the second equality holds because each free request incurs one unit
of waiting cost per unit of time. The following lemma is the
counterpart of Lemma~\ref{lem:ledger}. Its proof counts the waiting
times of both endpoints of each output pair.

\begin{lemma}
\label{lem:nb-ledger}
Non-bipartite VRM satisfies
\[
  \ALG\le\frac{2\gamma}{\gamma-1}W_R.
\]
\end{lemma}
\begin{proof}
As in the proof of Lemma~\ref{lem:ledger}, let $P_i$ be the path used
in the $i$th augmentation, let $M_i$ be the offline matching right
before it, and write $D_i^+=D(P_i\setminus M_i)$,
$D_i^-=D(P_i\cap M_i)$, and $\varphi_i=\gamma D_i^+-D_i^-$.
Let $u_i$ be the request that initiates this augmentation, let $v_i$ be
the other endpoint of $P_i$, and let $t_i$ be the time of this
augmentation. Define
\[
  S=\sum_{i=1}^m(t_i-a_{u_i}),\qquad
  T=\sum_{i=1}^m(t_i-a_{v_i}).
\]

By Eq.~\eqref{eq:nb-ready}, $\varphi_i\le\gamma(t_i-a_{u_i})$ for every
$i$, so $\sum_i\varphi_i\le\gamma S$. Because the $i$th augmentation
changes the total TA distance of $\Moff$ by $D_i^+-D_i^-$, the sum
$\sum_i(D_i^+-D_i^-)$ is the total TA distance of the final offline
matching, so it is nonnegative. Because
$(\gamma-1)(D_i^++D_i^-)+(\gamma+1)(D_i^+-D_i^-)=2\varphi_i$ for every $i$, we
have
\[
  \sum_{i=1}^m(D_i^++D_i^-)
   =\frac{2}{\gamma-1}\sum_{i=1}^m\varphi_i
       -\frac{\gamma+1}{\gamma-1}\sum_{i=1}^m(D_i^+-D_i^-)
   \le\frac{2\gamma}{\gamma-1}S.
\]

By the triangle inequality and $t_i\ge\max\{a_{u_i},a_{v_i}\}$,
the cost of output pair $u_iv_i$ is
\begin{align*}
  d(x_{u_i},x_{v_i})+(t_i-a_{u_i})+(t_i-a_{v_i})
   &=D(u_i,v_i)+2\bigl(t_i-\max\{a_{u_i},a_{v_i}\}\bigr)\\
   &\le D_i^++D_i^-+2(t_i-a_{v_i}).
\end{align*}
Summing over the pairs gives
\[
  \ALG\le\frac{2\gamma}{\gamma-1}S+2T
    \le\frac{2\gamma}{\gamma-1}(S+T)
    =\frac{2\gamma}{\gamma-1}W_R,
\]
where the second inequality uses $\gamma>1$ and the equality uses
$S+T=W_R$.
\end{proof}

\subsection{Charging the Waiting Time to the Optimal Matching}
\label{sec:nb-charging}

We now bound $W_R$ by adapting the charging scheme in
Sections~\ref{sec:charging}--\ref{sec:case2}. Fix a minimum-cost perfect
matching $M^*$ on $R$ under $D$, so that $D(M^*)=\OPT$, and write $M_t$
for the offline matching at time $t$. Consider a free request $u$ at time
$t$. Because $u$ is not saturated by $M_t$ and $M^*$ is perfect, the
component of $M_t\symdiff M^*$ containing $u$ is an $M_t$-augmenting
path from $u$ to another request. We call this path the \emph{optimal path}
of $u$ at time $t$, denote it by $P^*_u(t)$, and call its last vertex the
\emph{terminal request}. The terminal request is not saturated by $M_t$, so it
is either free or not yet arrived.

\begin{lemma}
\label{lem:nb-structure}
Let $u$ and $w$ be distinct free requests at time $t$. If $w$ is the
terminal request of $P^*_u(t)$, then $P^*_u(t)$ and $P^*_w(t)$ have the same
edges. Otherwise, $P^*_u(t)$ and $P^*_w(t)$ share no edge.
Moreover, every interior vertex of an optimal path has arrived.
\end{lemma}
\begin{proof}
The paths $P^*_u(t)$ and $P^*_w(t)$ are the components of $M_t\symdiff M^*$
containing $u$ and $w$, respectively. If $w$ is the terminal request of
$P^*_u(t)$, the two components coincide, so the two paths have the same
edges. Otherwise, $w$ is neither $u$ nor the terminal request
of $P^*_u(t)$. Because $w$ is not saturated by $M_t$, it is not an interior
vertex of $P^*_u(t)$ either. Hence $w$ is not on $P^*_u(t)$, and the two
components are different and share no edge. Next, an optimal path
alternates between edges of $M^*$ and edges of $M_t$, so each of its
interior vertices is incident to an edge of $M_t$. Because $M_t$ is a
matching on the requests that have arrived, every interior vertex has
arrived.
\end{proof}

The following lemma is the counterpart of Lemma~\ref{lem:window}.

\begin{lemma}
\label{lem:nb-window}
Fix a stable time $t$ and a free request $u$ such that the terminal request of
$P^*=P^*_u(t)$ has arrived, and let $D^*=D(P^*\cap M^*)$. Then
\[
  D^*\ge\age(u,t)
  \qquad\text{and}\qquad
  D^*\ge\frac{\age(z,t)}{\gamma+1}\ \text{ for every vertex $z$ of $P^*$.}
\]
\end{lemma}
\begin{proof}
Let $D^-=D(P^*\cap M_t)$, so that $D(P^*)=D^*+D^-$. By
Lemma~\ref{lem:nb-structure} and the assumption, all vertices of $P^*$ have
arrived. The terminal request of $P^*$ is not saturated by $M_t$, so it is
free. Thus, $P^*$ is an $M_t$-augmenting path between two free requests, and
Eq.~\eqref{eq:nb-stable} gives
$\gamma D^*-D^-=\varphi_{M_t}(P^*)>\gamma\age(u,t)$. This is the inequality
that the proof of Lemma~\ref{lem:window} obtains from
Lemma~\ref{lem:stable}. The rest of that proof then applies verbatim, with
$u$ and $z$ in place of $r$ and $v$.
\end{proof}

Because there are only finitely many event times, the waiting cost
incurred at them is zero. Thus, it suffices to charge the waiting cost at stable
times. We consider a stable time $t$ and a free request
$u$, and we charge the waiting cost of $u$ as follows.

\paragraph{Case 1: the terminal request of $P^*_u(t)$ has not arrived.}
We charge the waiting cost of $u$ to the last edge $e=pv$ of $P^*_u(t)$,
where $v$ is the terminal request. This edge belongs to $M^*$. By
Lemma~\ref{lem:nb-structure}, $p$ has arrived, since it is $u$ or an
interior vertex of $P^*_u(t)$. Because $v$ has not arrived, it is not
free. Thus, by Lemma~\ref{lem:nb-structure}, no other free request has an
optimal path containing $e$, and $e$ is charged at rate at most one.
Let $W_1$ be the total waiting cost charged in this case over all requests
and over time.

Fix an edge $e=pv\in M^*$, where $a_p\le a_v$. Whenever $e$ is charged
in this case at time $t$, we have $a_p\le t<a_v$. Because the charge
rate is at most one, the total charge to $e$ in this case is at most
$a_v-a_p\le D(p,v)$. Summing over all edges of $M^*$ gives
$W_1\le\sum_{pv\in M^*}D(p,v)=\OPT$.

\paragraph{Case 2: the terminal request of $P^*_u(t)$ has arrived.}
Let $D^*=D(P^*_u(t)\cap M^*)$ be the total TA distance of the edges of
$M^*$ on $P^*_u(t)$. For each edge $e\in P^*_u(t)\cap M^*$, we charge the
waiting cost of $u$ to $e$ at rate $D(e)/D^*$. Lemma~\ref{lem:nb-window}
gives $D^*\ge\age(u,t)$, and $\age(u,t)>0$ because no request arrives at
the stable time $t$. Thus, the rate is well defined. For each edge $e\in M^*$, let $q_e(t)$ be the sum of
the charge rates assigned to $e$ in this case by all free requests at a
stable time $t$, with $q_e(t)=0$ at event times.

Fix an edge $e\in M^*$. Choose one endpoint $z$ of $e$, and let
$b_e=a_z$. Consider a stable
time $t$ with $q_e(t)>0$. Let $u_e(t)$ be a request charging $e$ at time
$t$, let $P^*_e(t)=P^*_{u_e(t)}(t)$ be its optimal path, and let
$D^*_e(t)=D(P^*_e(t)\cap M^*)$. By Lemma~\ref{lem:nb-structure}, the
terminal request of $P^*_e(t)$ is the only other request that can charge
$e$ at time $t$, and its optimal path has the same edges as $P^*_e(t)$.
Thus, at most two requests charge $e$, and each of them charges $e$ at rate
$D(e)/D^*_e(t)$. Hence $q_e(t)\le2D(e)/D^*_e(t)$.

The waiting cost of each of these requests increases at rate one and is
split among the edges of $P^*_e(t)\cap M^*$. Thus, $q_e(t)\le2$.
Moreover, $z$ is a vertex of $P^*_e(t)$, and in Case~2 every vertex of
$P^*_e(t)$ has arrived. Thus, $z$ has arrived. Because no request arrives
at the stable time $t$, we have $\age(z,t)=t-b_e>0$. The second bound of
Lemma~\ref{lem:nb-window}, applied to $u_e(t)$, gives
$D^*_e(t)\ge(t-b_e)/(\gamma+1)$.
Because $D^*_e(t)\le D(M^*)=\OPT$, this bound gives
$t\le b_e+(\gamma+1)\OPT$. As in the proof of Lemma~\ref{lem:edge}, we
have
\begin{equation}
  \begin{aligned}
    q_e(t)&\le\min\Bigl\{2,\,\frac{2(\gamma+1)D(e)}{t-b_e}\Bigr\}
      &&\text{if } t\in(b_e,\,b_e+(\gamma+1)\OPT],\\
    q_e(t)&=0 &&\text{otherwise}.
  \end{aligned}
  \label{eq:nb-envelope}
\end{equation}
Integrating as in Eq.~\eqref{eq:integral}, we
have
\[
  \int_0^\infty q_e(t)\,dt
   \le2(\gamma+1)\Bigl(\frac{\OPT}{m}+D(e)\ln m\Bigr).
\]

\subsection{Proof of Theorem~\ref{thm:nonbipartite}}
\label{sec:nb-proof}

Adding the
charge $W_1$ in Case~1 and the charge to all edges in Case~2, we obtain
\begin{align*}
  W_R&=W_1+\sum_{e\in M^*}\int_0^\infty q_e(t)\,dt\\
     &\le\OPT+\sum_{e\in M^*}2(\gamma+1)\Bigl(\frac{\OPT}{m}+D(e)\ln m\Bigr)\\
     &=\bigl[1+2(\gamma+1)(1+\ln m)\bigr]\OPT.
\end{align*}
The last equality holds because $M^*$ has $m$ edges and
$\sum_{e\in M^*}D(e)=\OPT$.
Combining this bound with Lemma~\ref{lem:nb-ledger} gives
\[
  \ALG\le\frac{2\gamma}{\gamma-1}
     \bigl[1+2(\gamma+1)(1+\ln m)\bigr]\OPT.
\]
For $\gamma=2$, the coefficient is
$4[1+6(1+\ln m)]=28+24\ln m$. Finally, the algorithm does not need
to know the metric space or $m$ in advance, since $m$, $M^*$, and $\OPT$ are used only in the analysis.

\section{Concluding Remarks}
\label{sec:conclusion}

Several ingredients of the analysis are shared with previous analyses of
RM and VRM. The bound on the total length of the augmenting paths in
Lemma~\ref{lem:ledger} is from \cite{raghvendra2016robust,kuo2025online},
and the optimal-path
construction in Section~\ref{sec:charging} and the path-length comparison
in the proof of Lemma~\ref{lem:window} are
from~\cite{nayyar2017input,kuo2025online}.
The distinguishing feature of our analysis is that it charges waiting
time over time rather than net-cost per augmentation. In particular, it
does not use the geometric decomposition of the metric space that
underlies the analyses in~\cite{nayyar2017input,kuo2025online}, and it
does not use the
monotonicity of the minimum net-cost under
augmentations~\cite[Lemma~4.2]{kuo2025online}.

Without delays, no deterministic algorithm is polylogarithmic-competitive
in high dimensions. In $d$-dimensional Euclidean space, every
deterministic algorithm for the MBPM problem has competitive ratio
$\Omega(m^{1-1/d})$~\cite{raghvendra2018optimal,nayyar2017input}. In
contrast, VRM, which is built on RM, is $O(\log m)$-competitive for the
MBPMD problem in any metric space. From this perspective, allowing delays makes matching easier for
deterministic algorithms, although the two problems are not directly
comparable.

\clearpage
\pdfbookmark[1]{AI Usage and Methodology}{sec:ai-usage}
\section*{AI Usage and Methodology}

\paragraph{Background.}
After our earlier paper on VRM~\cite{kuo2025online}, we did not expect
VRM to be polylogarithmic-competitive, because RM, on which it is based,
is not polylogarithmic-competitive in high dimensions (see
Section~\ref{sec:conclusion}). We therefore did not plan to analyze VRM
further. The deterministic $O(\log^5 m)$-competitive algorithm of Dufay
and Wattenhofer for the MPMD problem~\cite{dufay2026deterministic}
motivated us to revisit the MBPMD problem.
In October 2025, we tried to design a new algorithm for the MBPMD problem
with GPT-5 (OpenAI) and substantial human input, but these attempts
failed.

\paragraph{How the result was found.}
In September 2026, we first gave Claude Fable~5.1 (Anthropic) a single prompt
that included our paper on VRM~\cite{kuo2025online}. In this prompt, we
asked the model to read that paper and then to consider the same
problem, the MBPMD problem on the line, and either to design a
deterministic polylogarithmic-competitive online algorithm or to give a
lower bound showing that no such algorithm exists. This attempt
failed. We then gave the same prompt to GPT-6 Astra (OpenAI) in the same
working directory, which still contained the notes and code of that
attempt. We expected that, if the model succeeded, it would design a new
algorithm.

\paragraph{What GPT-6 Astra contributed.}
Instead, GPT-6 Astra proved that VRM is $O(\log m)$-competitive
in any metric space, although the problem was posed on the line. There was
almost no human input. We repeatedly asked the model to continue
until the proof was complete. The core of its proof is the
charging argument now in Sections~\ref{sec:charging}--\ref{sec:case2}. In a separate session, we asked the same
model to extend the result to the MPMD problem, again with almost no
human input. The extension in Section~\ref{sec:nonbipartite} comes from
that session.

We found afterwards that GPT-6 Astra's notes cite the notes of the
Claude Fable~5.1 attempt and that its scripts reuse the code of that
attempt. Claude Fable~5.1's notes already distinguished the two cases of
Section~\ref{sec:charging}. They contained a charging scheme for Case~1
but none that bounded the waiting cost in Case~2.

\paragraph{Verification and presentation.}
We verified the proofs found by GPT-6 Astra by rewriting them with Claude
Opus~5.5 (Anthropic), which also helped us simplify and present the
results and edit the text. Claude Fable~5.1 also checked an early
version of Section~\ref{sec:nonbipartite} and revised its notation. We
checked every step of the rewritten proofs ourselves. This
paper presents the rewritten proofs, which include one slight
improvement. The next paragraph describes it.

We first explained to Claude Opus~5.5 our own interpretation of the
original proof for the MBPMD problem found by GPT-6 Astra and then discussed how one might
arrive at this proof. After this discussion, Claude Opus~5.5 noticed
that, in the proof of Lemma~\ref{lem:edge}, the charge rate $q_e(t)$ can
also be bounded by~$1$ (see Eq.~\eqref{eq:envelope}). This observation
simplifies both original proofs found by GPT-6 Astra and improves the
resulting competitive ratios.

While rewriting the proofs, we found another improvement, which this
paper does not use. Using MV servers and~\cite[Lemma~4.1]{kuo2025online},
we can show that Case~1 is not needed. For $\gamma=2$, this improves the
ratio in Theorem~\ref{thm:main} from $24+18\ln m$ to $18+18\ln m$.
We later discovered that the original proof found by GPT-6 Astra already
contains this argument.

However, because the proof with Case~1 does not use MV servers, it is
simpler. Moreover, Section~\ref{sec:nonbipartite} adapts it to the MPMD
problem, whereas the other improvement relies
on~\cite[Lemma~4.1]{kuo2025online}, which does not extend to the MPMD
problem. For these reasons, we present the proof with Case~1.

\bibliographystyle{alphaurl}
\bibliography{refs}

\appendix
\section{Validity of the Algorithm}
\label{app:validity}

In this appendix, we prove Lemma~\ref{lem:cycles}. The proof is similar
to the corresponding proofs for RM~\cite{raghvendra2016robust} and
VRM~\cite{kuo2025online}. In particular, it uses the relaxed dual
feasibility and matched-edge tightness conditions
of~\cite{raghvendra2016robust}
and~\cite[Section~3.2.1, Invariants~(I1) and~(I4)]{kuo2025online}.

\begin{proof}[Proof of Lemma~\ref{lem:cycles}]
We prove by induction that $H$ has no negative cycle.
Initially, $H$ has no cycle. When a request arrives, it is free and
has no incoming edge. Similarly, a newly arrived server has no outgoing
edge. Thus, newly arrived vertices cannot create a directed cycle.

Next, consider an augmentation and assume that there is no negative cycle
before the augmentation. We first work with the state right before the
augmentation. Let $T$ be the nonempty set of free servers.
Every request has a directed edge to a server in $T$, and every matched
server has a directed edge to the request matched to it. Thus, every
arrived vertex is in $T$ or has a directed path to $T$. For each arrived vertex $v$, let $\ell(v)$
be its shortest distance to $T$ in $H$. Because there is no negative cycle, this
distance is finite and is attained by a simple path. Also, $\ell(s)=0$ for
every $s\in T$.

To show that the augmentation does not create a negative cycle, we assign
a real-valued variable to each arrived vertex as follows:
\[
  z_r=\ell(r)\quad(r\in R),\qquad
  z_s=-\ell(s)\quad(s\in S).
\]
Thus, $z_r+z_s=\ell(r)-\ell(s)$ for every edge $rs$ at this state.

For any unmatched edge $rs$, the shortest-distance inequality gives
\begin{equation}
  z_r+z_s\le\gamma D(r,s).
  \label{eq:dual-upper}
\end{equation}
For any matched edge $rs$, the only outgoing edge of $s$ is the directed
edge from $s$ to $r$, so $\ell(s)=-D(r,s)+\ell(r)$. Thus,
\begin{equation}
  z_r+z_s=D(r,s)\le\gamma D(r,s).
  \label{eq:dual-tight}
\end{equation}
Hence, Eq.~\eqref{eq:dual-upper} holds for every edge.

Let $P$ be the minimum augmenting path used in the augmentation. Because
$P$ is a shortest path from its first vertex to $T$ in $H$, every
directed edge of $P$ from $u$ to $v$ satisfies $\ell(u)=w+\ell(v)$,
where $w$ is the weight of this edge in $H$. A forward edge $rs$ of $P$
is directed from $r$ to $s$ and has weight $\gamma D(r,s)$. Thus, it
satisfies $z_r+z_s=\ell(r)-\ell(s)=\gamma D(r,s)$. These forward edges are exactly the edges that
become matched in this augmentation.

After the augmentation, each request $r$ on $P$ is matched by the forward
edge $rs$ of $P$ that leaves $r$. We decrease $z_r$ by $(\gamma-1)D(r,s)$,
and all other variables remain unchanged. Because only request variables
decrease, Eq.~\eqref{eq:dual-upper} continues to hold for every edge.
Each newly matched edge $rs$ had $z_r+z_s=\gamma D(r,s)$ before this
update, so it now has $z_r+z_s=\gamma D(r,s)-(\gamma-1)D(r,s)=D(r,s)$,
and Eq.~\eqref{eq:dual-tight} holds for it.
For an old matched edge that remains in the matching, neither endpoint is
on $P$, so Eq.~\eqref{eq:dual-tight} also continues to hold.

Consider any cycle $C$ in $H$ after the augmentation, and let $C^+$ and
$C^-$ be the sets of unmatched and matched edges of $C$, respectively.
The weight of $C$ is $\gamma D(C^+)-D(C^-)$. By Eq.~\eqref{eq:dual-upper}
for the edges in $C^+$ and Eq.~\eqref{eq:dual-tight} for the edges in
$C^-$, we have
\[
  \gamma D(C^+)-D(C^-)
  \ge\sum_{rs\in C^+}(z_r+z_s)-\sum_{rs\in C^-}(z_r+z_s)=0.
\]
The last equality holds because unmatched and matched edges alternate
along $C$, so every vertex of $C$ is incident to exactly one edge in $C^+$
and exactly one edge in $C^-$. Thus, $C$ has nonnegative weight, which
completes the induction.
\end{proof}

\end{document}